\documentstyle[psfig,conf_iap,10pt]{article}
\begin{document}
\def\Sec{${}^{\prime\prime}$}
\heading{%
%
The MUNICs project -- a multicolor survey of distant galaxies \\
%
} 
\par\medskip\noindent
\author{%
C. Mendes de Oliveira$^{1,2}$, N. Drory$^{2}$, U. Hopp$^{2}$, R. Bender
$^{2}$, R.P. Saglia$^{2}$
}

\address{%
Instituto Astron\^omico e Geof\'{\i}sico, Av Miguel St\'efano 4200,
04301-904, S\~ao Paulo, Brazil
}
\address{%
Universit\"ats-Sternwarte, Ludwig-Maximilians-Universit\"at,
Scheinersstrasse 1, 81679 Munich, Germany
}
%

\begin{abstract}
The MUNICS project is an ongoing imaging survey designed to cover 3
sq.  degrees in V,R,I,J,K$^{\prime}$.  We describe here partial results
of the project concerning the clustering properties of K$^{\prime}$ $<$
19.5 galaxies in scales of 3.6$^{\prime\prime}$ to 63.0$^{\prime\prime}$ 
over an area of $\sim$
800 arcmin$^2$.  We present K$^\prime$ data for a sample of 20 fields,
five of which contain $z > 0.5$ radio-loud quasars with steep spectra,
eight contain $z > 0.5$ radio-loud quasars with flat spectra and seven
are high-galactic latitude fields with no quasars in them.  The
two-point angular correlation function for the total sample shows
significant clustering at $\sim$ 5$\sigma$ level of K$^\prime$=19.5
galaxies. The correlation angle of the galaxies is $\theta_0$ = 1.7$^{\prime\prime}$
$\pm$ 0.4$^{\prime\prime}$ for K$^\prime$ $<$ 19 mag and $\theta_0$ = 1.0$^{\prime\prime}$ $\pm$ 0.2$^{\prime\prime}$
for K$^\prime$ $<$ 19.5 mag.  When the correlation functions for the
subsamples are considered, the mean $\omega$($\theta$) amplitude of the
fields which contain steep-spectra $z > 0.5$ radio-loud quasars is determined to
be $\sim$ 2.0 -- 2.5 that of the high-galactic latitude fields.

\end{abstract}

\section{Introduction} %

  The MUNICs project has four main goals:  1) determine the space
density of $z$ $\sim$ 1 clusters; 2) measure large-scale structure at
redshifts $z >$ 0.5; 3) test the number density evolution of elliptical
galaxies; 4) select a sample of large--$z$ elliptical galaxies for
populations studies.  The basic observational set-up and some
preliminary results were described in \cite{Men}.

   We describe below the preliminary results on the clustering
properties of K$^{\prime}$ $\sim$ 19.5 mag  galaxies for three
subsamples:  1) seven ``empty'' high-galactic latitude fields, 2) five
fields containing $z > $ 0.5 radio-loud quasars with steep-spectra,
i.e., 
with spectral indices (as derived from the ratio of the fluxes
at the 11cm- and 6cm-bands), of $\alpha$ $>$ --0.6 and 3) eight fields
containing $z > $ 0.5 radio-loud quasars with flat spectra, with $\alpha$ $
< $ --0.6.  The quasars were chosen from \cite{Ver}
and the criteria used for their selection
were that they have $z >$ 0.5, their galactic latitude be $b$ $>$
40 degrees, to avoid heavy star contamination and they be detected
at 6 cm and 11 cm.

  The K$^{\prime}$ images used in this study were taken at the
3.5m-tel at Calar Alto with the Omega camera.  The typical usable
area for each field was 40 arcmin$^2$.

\section{The angular correlation function in the K$^{\prime}$ band}

  A measurement of the degree of galaxy clustering in our fields can be
made by determining their projected 2-point
correlation function, i.e.,
the angular 2-point correlation function, $\omega$($\theta$).  This is
generally fit by a power law of slope --0.8 and amplitude given by
integration of Limber's equation for the 3-d 2-point correlation
function over the galaxy redshift distribution.

  We calculate the galaxy angular correlation function for the three
subsamples and the complete sample of 20 fields using the formula
$\omega$($\theta$)=(DD--2DR$+$RR)/RR \cite{Lan}, where 
for a given angle separation DD is the number of galaxy pairs
in our data sample, DR is the
number of pairs between the data and a uniform random sample and
RR is the number of random pairs. We calculate $\omega$($\theta$)
between 3.6$^{\prime\prime}$ 
and 63.0$^{\prime\prime}$ in log bins of width 0.25.

  The fitted $\omega$($\theta$) amplitudes at $\theta$ = 1
degree, assuming a $\theta^{-0.8}$ power-law for the
radio/steep-source, radio/flat-source and empty fields and for the
combined sample are given in Table 1. The corresponding plots for the
three subsamples are shown in Figs. 1 and 2.  The correlation function
was determined for each field individually and the plotted values are
the averages over the fields. The error bars indicate the rms scatter
between the fields. The 1$\sigma$ error in the mean is comparable to
the symbol sizes.
To account for the field sizes we calculate an integral constraint (IC)
assuming that the form of $\omega$($\theta$) is given by a power law
with a slope of --0.8. The solid lines in Figs. 1 and 2 correspond to
the best least-squares fits to the function A($\theta^{-0.8}$ -- 13.5)
where the value for the amplitude A is given in Table 1 and the average
IC is 13.5 $\pm$ 1. Only
angular separations for which the IC 
is $<$ 50\% the raw correlation function are considered.
The error on A was estimated from the field-to-field scatter 
derived by fitting $\omega$($\theta$) of the individual
fields.

\begin{figure}
\centerline{\vbox{
\psfig{figure=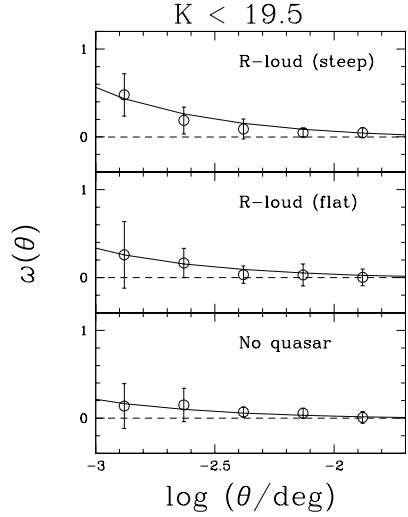,height=7.cm}
}}
\caption[]{The angular correlation function, $\omega$($\theta$), of the
K$^{\prime}$ $<$ 19.5 mag galaxies on five fields containing radio-loud
steep-spectrum sources, eight with radio loud flat-spectrum sources and
seven empty fields.  The solid lines correspond to the best fits to the
function A($\theta^{-0.8}$ -- 13.5) where the amplitudes at $\theta$ =
1 degree (A) are listed in Table 1 and the integral constraint is 13.5
$\pm$ 1.  The error bars indicate the rms scatter
between the fields.  }
\end{figure}

\begin{figure}
\centerline{\vbox{
\psfig{figure=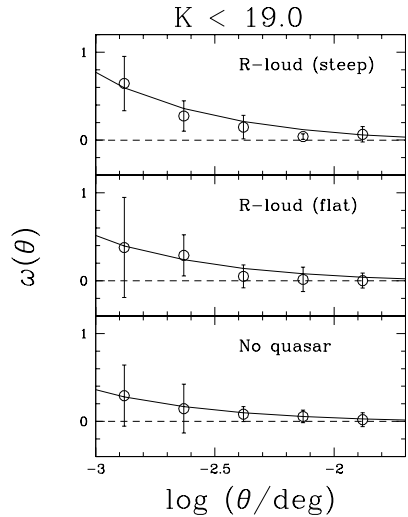,height=7.cm}
}}
\caption[]{Same as Fig. 1, but for K$^{\prime}$ $<$ 19.0 mag galaxies.
}
\end{figure}

\section {Results}

  The correlation function of galaxies in the K$^{\prime}$ band as a
function of limiting magnitude is not well determined mainly due to the
small sizes ( $<$ 2$^{\prime}$ -- 3$^{\prime}$ on a side) of the NIR detectors which have
not allowed wide-field K$^{\prime}$ imaging surveys to be performed. This will
certainly change with the several wide-field ongoing or planned surveys
(see this conference).

   We summarize in Table 2 the results of the four studies (including
ours) which have derived the K$^{\prime}$ band correlation functions
to date. We list amplitudes normalized to $\theta$ = 1 degree and
their 1$\sigma$ errors.
As can be seen in Table 2, the $\omega$($\theta$) amplitude
derived in this study from 800 arcmin$^2$
is similar to that estimated by \cite{Roc}
for 17 fields covering a total of 101.5 arcmin$^2$,
each containing a radio-galaxy at $z \sim$
1.1,  for a similar magnitude limit.  Our results are within the
values expected from models of stable clustering with $\epsilon$ = 0
\cite{Roc}.
Further discussion of these results is
deferred to a later paper.

  We also find a marginally significant 2$\sigma$ 
enhancement on the clustering amplitude by a factor of 2.0 -- 2.5 for
fields with steep-spectra radio-loud quasars, as compared to the fields
with no quasars (see Table 1).
This result has, however, a weak significance and
must await confirmation for a larger sample. If this is correct it
could be interpreted as 
an extension of the effect already observed for quasars at
lower redshifts.

\begin{center}
\begin{tabular}{l r r c}
\multicolumn{3}{l}{{\bf Table 1.} Amplitudes of the $\omega$($\theta$)
at $\theta$ = 1 degree}
\\
\hline
\\
\multicolumn{1}{c}{ }&\multicolumn{1}{c}{K$^{\prime} <$ 19.5}&
\multicolumn{1}{c}{K$^{\prime} <$ 19.0} & \multicolumn{1}{c}{area
(arcm$^2$)
} 
\\
\hline
\\
Radio steep & 23.0$\pm$6.4 $\times$ 10$^{-4}$  & 31.5$\pm$10.2 $\times$
10$^{-4}$  & 200  \\
Radio flat  & 13.7$\pm$5.5 $\times$ 10$^{-4}$  & 21.0$\pm$9.5 $\times$
10$^{-4}$  & 320  \\
Empty fields & 8.8$\pm$5.1 $\times$ 10$^{-4}$   & 14.8$\pm$7.7 $\times$
10$^{-4}$  & 280  \\
All fields  & 14.3$\pm$3.2 $\times$ 10$^{-4}$ & 21.4$\pm$4.7 $\times$
10$^{-4}$  & 800  \\
\\
\hline
\end{tabular}
\end{center}

\begin{center}
\begin{tabular}{l c c c}
\multicolumn{3}{l}{{\bf Table 2.} K$^{\prime}$ angular correlation functions
derived to date } \\
\hline
\\
\multicolumn{1}{c}{K$^{\prime}$ limit } &
\multicolumn{1}{c}{Amplitude ($\times$ 10$^{-4}$) } &
\multicolumn{1}{c}{Environment} & 
\multicolumn{1}{c}{Reference}
\\
\hline
\\
K $<$ 15 &  430$\pm$20
& field & \cite{Bau} \\
K $<$ 15 &  350$\pm$20 
& field & "                 \\
15 $<$ K $<$ 16 &  96$\pm$5
& field & "  \\
15 $<$ K $<$ 16 &  84$\pm$7
& field & "  \\
K $<$ 19 &  21.4$\pm$4.7 
& field+quasar & this work \\
K $<$ 19.5 & 14.3$\pm$3.2 
& field+quasar & " \\
K $<$ 19.5 & 13.5$\pm$3.2 
& radio galaxy & \cite{Roc} \\
K $<$ 20.0 & 13.3$\pm$2.3 
& radio galaxy & " \\
K $<$ 20.5 & 8.1$\pm$1.6 
& radio galaxy & " \\
K $<$ 21.5 & 11.4$\pm$1.6 
& field & \cite{Car} \\
\\
\hline
\end{tabular}
\end{center}

\acknowledgements{
We thank the staff at Calar Alto for helping with the observations.
CMdO acknowledges the financial support from the Alexander von Humboldt
Foundation. This work was supported by the Sonderforschungsbereich
SFB375.}

\vfill
\end{document}